\documentclass[twocolumn,preprintnumbers,showpacs,prl]{revtex4}

\usepackage{graphicx}
\usepackage{amsfonts}
\usepackage{amsmath}
\usepackage{bm}

\def\beq{\begin{equation}}
\def\eeq{\end{equation}}
\def\bea{\begin{eqnarray}}
\def\eea{\end{eqnarray}}

\newcommand{\vth}{\ensuremath{v_{th}}}

\newcommand{\nuhc}{\ensuremath{\nu_{ei}^{hc}}}
\newcommand{\keps}{\ensuremath{{\bm k}\!\cdot{\bm \epsilon}}}

\newcommand{\Real}[1]{\ensuremath{\mathfrak{\rm Re\!}\left\{#1\right\}}}

\begin{document}


\title{Nonlinear collisional absorption of laser light
       in dense, strongly coupled plasmas}

\author{A.\ Grinenko}
\author{D.O.\ Gericke}
\affiliation{Centre for Fusion, Space and Astrophysics,
             Department of Physics, University of Warwick,
             Coventry CV4 7AL, United Kingdom}

\date{\today}

\begin{abstract}
We present a new theoretical approach for collisional absorption
of laser energy in dense plasmas which accommodates arbitrary
frequencies and high intensities of the laser
field. We establish a connection between laser absorption by
inverse Bremstrahlung and the stopping power. This relation is
then applied to include strong correlations beyond the mean field
approach. The results show an excellent agreement with molecular
dynamics simulations up to very high coupling strength.
\end{abstract}

\maketitle

Understanding the interaction of intense radiation with strongly coupled
plasmas is crucial for the design and critical evaluation of targets for
inertial confinement fusion. To accommodate the symmetry conditions,
the absorption of laser energy must be carefully determined starting from the
early stages \cite{Lindl_04,MDWW_09}. Similar data are required for
fast ignition by ultra-intense lasers due to creation of a dense plasma by
the nanosecond pre-pulse \cite{K_07}. Least understood are laser-plasma
interactions that involve nonideal and partially degenerate electrons with
$\Gamma \!=\! (e^2/k_B T_e) (4\pi n_e/3)^{1/3} \!\sim\! 1$ and
$n_e \Lambda_e^3 \!=\! n_e (2\pi\hbar^2/m_e k_B T_e)^{3/2} \!\sim\! 1$,
respectively. Such conditions are also created in contemporary warm dense
matter experiments \cite{KNCD_08,GGGV_08} and laser-cluster interactions
\cite{MHBK_08,BHS_07}.

The dominant absorption mechanism for lasers with intermediate intensities
typical for inertial confinement fusion is the inverse bremsstrahlung. A first
description of this process was presented by Dawson \& Oberman for weak fields
\cite{DO_62} which was later extended to arbitrary field strengths
\cite{DMDK_94}. Due to the classical description, these results become
inapplicable for dense, strongly coupled plasmas. This break-down is avoided by
a rigorous quantum kinetic description applying the Green's function formalism
\cite{KBBS_99,BSHK_01} or the quantum Vlasov equation \cite{KP_01}. For weak
laser fields that allow for a formulation within linear response, strong
electron-ion collisions were included into a quantum description
\cite{RRRW_00,WMRR_01}.

All approaches mentioned above are formulated in the high-frequency limit which
requires the number of electron-ion collisions per laser cycle to be small.
In this limit, the electron-ion interaction has a collective rather than 
a binary character and the laser energy is coupled into the plasmas via the
induced polarization current. On the other hand, binary collisions dominate
laser absorption in the low-frequency limit and a Drude-like formulation
follows. At intermediate frequencies, both strong binary collisions and
collective phenomena have to be considered simultaneously. Interestingly, such
conditions occur for moderate heating at the critical density of common
Ny:Yag lasers. The strong restrictions on the applicability of the theories
above are also related to the assumption of nearly equilibrium electron
distributions either in the reference frame of the ions for low frequencies
or in the frame of freely moving electrons for high frequencies of the laser.

In this Letter, we present a description of collisional absorption that bridges
between the high- and low-frequency limits and incorporates weak collective
interactions and strong binary collisions simultaneously. To this end, we 
formally split the interactions into a weak interaction part and additional
contributions due to hard collisions. The latter constitutes a friction
between the electron and ion fluids and can be treated as the stopping power
of ions in the electron fluid. Thus, two basic energy absorption mechanism, the
stopping power and collisional absorption, are connected which allows one to
apply well developed models for the stopping power (see, e.g.,
Refs.~\cite{GS_99,G_02}) to the problem of laser absorption in plasmas.
Restrictions with respect to the laser frequency are avoided by expanding the
electrons distribution in a generalized Kramers-Hennenberger frame that follows
the center of the electron fluid and is determined by the driving field and the
friction between the two species. The description of the collective electron
response can be kept almost unchanged from earlier approaches 
\cite{DMDK_94,KP_01}. Due to the use of quantum approaches for the stopping
power and a quantum dielectric function, no {\em ad hoc} cutoffs must be
introduced and the theory stays reliable for strong electron-ion interactions
and degenerate electrons. The few assumptions made are justified by the
unprecedented agreement with molecular dynamic (MD) simulations
\cite{PG_98,HSBK_05} up to very high coupling strengths.

Collisional absorption of laser energy is commonly characterized in terms of
a frequency-dependent electron-ion collision frequency defined as \cite{S_65}
\beq
\nu_{ei}(\omega) = \frac{4\pi \omega_0^2}{\omega_p^2}
                   \frac{\langle {\bm j} \cdot {\bm E} \rangle}
                        {\langle {\bm E} \cdot {\bm E} \rangle}
\;\triangleq\:     \frac{4\pi \omega_0^2}{\omega_p^2}\;
                   \Real{\sigma(\omega)} \,,
\label{eq:coll_freq_def}
\eeq
where the brackets denote the average over one period of the laser field with
frequency $\omega_0$ and $\omega_p \!=\! (e^2 n_e/m_e)^{1/2}$ is the plasma
frequency of the target electrons. The collision frequency $\nu_{ei}$ contains
the same information as the dynamic conductivity $\sigma$ and is determined by
the electron current $\bm j$. This current, in turn, is the first moment of the
electron distribution $f_e({\bm p},t)$. We, therefore, base our statistical
description on a general quantum kinetic equation for the electrons in
an external field that has, for the homogenous plasmas considered, the form
 \cite{KBBS_99}
\beq
\frac{\partial f_e({\bm p},t)}{\partial t}
 - e {\bm E}_{ext}(t) \cdot \frac{\partial f_e({\bm p},t)}{\partial {\bm p}}
 = I_{ee}({\bm p},t) + I_{ei}({\bm p},t) \,.
\label{eq:boltzmann}
\eeq
Here, we will assume harmonic external electric fields:
${\bm E}_{ext}(t) = {\bm E_0} \sin{(\omega_0 t)}$. The form of the collision
integrals on the r.h.s. is not specified at this moment.

The current is obtained by multiplying Eq.~\eqref{eq:boltzmann} by $e{\bm p}/m$
and integrating over the free momentum. The terms on the l.h.s.\
are easily evaluated. The electron-electron collision term vanishes. The
electron-ion collision integral $I_{ei}$ is formally split into a part given by
the dynamically screened first Born approximation and the remainder that
contains the corrections due to hard collisions. Using the fact that the
treatments based on the quantum Vlasov equation and the dynamic Born
approximation are equivalent (compare Refs.~\cite{BSHK_01,KP_01}), the weak
coupling part of the collision integral can be written as a self-consistent
polarization ${\bm P}$ associated with the Hartree term. We then obtain for the
electron current
\beq
\frac{d{\bm j}}{dt}
  = \frac{\omega_p^2}{4\pi}
       \bigl[ {\bm E_0} \sin{(\omega_0 t)}+ \langle {\bm P(t)} \rangle \bigr]
    + \frac{e}{m} \int \! d {\bm p} \; {\bm p} \; I^{hc}_{ei} \,,
\label{eq:mot_current}
\eeq
\noindent
where the collision integral $I^{hc}_{ei}$ contains the corrections due to
hard collisions only.

Let us now consider the last term associated with the hard collisions in more
detail. Since the electron quiver velocity is practically always much larger
than the ion thermal motion, the ions can be considered to move with a common
velocity $V$ with respect to the electrons. In such a situation, the form of
the last term in Eq.~\eqref{eq:mot_current} is by definition the stopping
power, i.e., the energy loss per unit length, for the ions in an electron gas
(see, e.g., Ref.~\cite{GS_99}). Spatial correlations in the ion component can
be modelled as a correlated beam \cite{A_78,ZD_97}. Since the last term
contains only the contribution of hard collisions, it can be calculated as the
difference of the full stopping power (including hard collisions) and the
stopping power calculated from the (quantum) Lenard-Balescu equation describing
the weak interactions. The dynamics of screening is of minor importance when
describing the hard collisions, particularly in the low velocity case, but
cannot always be neglected. For that reason, only the contribution of hard
collisions is treated in this average manner and the full dynamics is kept in
the leading, weak coupling term of the Born series.

It is now convenient to cast the effect of the stopping power associated with
the hard collisions as a friction force or a collision frequency
\beq
\frac{e}{m} \int \!  d {\bm p} \; {\bm p} \; I^{hc}_{ei}
 = \frac{e}{m} \frac{\partial \langle E_i \rangle}{\partial \bm x}(V)
 = \frac{e}{m} R(V) {\bm V}
 = \nuhc {\bm j} \,,
\eeq
where $\bm x$ and $\bm V$ are pointing in the same direction as the external
field $\bf E_0$ and $V$ is the magnitude of the time-dependent velocity between
the electron and ion fluids. Again, the stopping power and the generalized
friction coefficient $R$ contain the effect of hard collisions only.

In general, the above model must be solved numerically. However, if the
friction coefficient $R$ is not velocity-dependent as during the linear
increase of the stopping power at small $V$, one can proceed analytically. This
regime is determined by the opposing forces created by the external field and 
friction due to hard collisions. As the stopping power is linear in $V$ up to
roughly the thermal velocity of the electrons, we have
\beq
V_{max} = \frac{1+(\nuhc / \omega_0)}{1+(\nuhc / \omega_0)^2} \; v_0
 \lesssim \vth \,,
\eeq
with the thermal velocity $\vth \!=\! (k_B T_e / m_e)^{1/2}$ and the free
quiver velocity $v_0 \!=\! e E_0 / m \omega_0$. In the high frequency limit,
we recover $v_0 \!<\! \vth$ as the restriction for the field strength while
$v_0 \!<\! [1 \!+\! (\nuhc/\omega_0) ] \, \vth$ follows in the low-frequency,
i.e.~$\nuhc/\omega_0 \gg 1$, limit. Thus, the region where
$R \!=\! \text{const}$ can be applied is extended to significantly higher field
amplitudes for strongly coupled plasmas and low to intermediate frequencies.

Given that $V \!<\! V_{max}$, we can formally solve Eq.~\eqref{eq:mot_current}
\bea
{\bm j} (t)  &=& -\frac{\omega_p^2}{4\pi}\frac{\gamma {\bm E_0}}{\omega_0}
                  \left[ \cos{(\omega_0t)}
                      - \frac{\nuhc}{\omega_0} \sin{(\omega_0t)} \right]
\nonumber \\
             &~& + \frac{\omega_p^2}{4\pi}
                   \int\limits_{-\infty}^{t}
                   \langle{\bm P}(\tau)\rangle \,
                   \exp(\nuhc [\tau-t]) \; d\tau \,,
\label{eq:current}
\eea
where we introduced $\gamma = [1+ (\nuhc/\omega_o)^2]^{-1}$. The total current
consists of contributions from the moving carriers and the polarization. The
strong collisions result in the second term in the brackets which is in phase
with the driver field. If the contribution from the polarization can be
neglected as in the low-frequency limit, Eq.~\eqref{eq:mot_current} gives the
well-known Drude formula for the conductivity:
$\sigma_D(\omega_0) \!=\! \omega_p^2 / 4\pi (\nuhc- i \omega_0)$.

By inserting Eq.~\eqref{eq:current} into the definition of the collision
frequency \eqref{eq:coll_freq_def}, one obtains
\beq
\nu_{ei}(\omega) = \gamma\nuhc + \frac{2\omega_0^2}{E_0^2}
                   \left\langle {\bm E}_{ext}(t)
                   \int\limits_{-\infty}^{t} \langle{\bm P}(\tau)\rangle \,
                   {\rm e}^{\nuhc(\tau - t)} d\tau \right \rangle \,.
\label{eq:frequency}
\eeq
The contribution of the hard collisions in the first term has the form of
the real part of the Drude conductivity.

Deriving an explicit expression for the second term in Eq.~\eqref{eq:frequency}
associated with the polarization is mainly straightforward. Differences to
earlier works arise from the additional hard collisions which define 
a generalized frame for the center of mass of the electron fluid. Here, we only
sketch the derivation using classical arguments; the quantum version of the
derivation follows the lines of Ref.~\cite{KP_01} and will be published
elsewhere \cite{GG_09b}.

To find the polarization field, one uses its relation to the self-consistent
potential of the plasma: ${\bm P} = -\nabla \Phi$. The field is, in turn,
determined by the Poisson equation
\beq
\nabla^2 \Phi = 4\pi n_e e \int d {\bm p} \, f_e({\bm p})
                - 4\pi Ze \sum_i \delta({\bm r - r}_i) \,,
\label{eq:Poisson}
\eeq
where we consider point-like ions positioned at ${\bm r}_i$. A direct solution
like for the Vlasov equation is not possible here as the collision integral
depends on the momentum ${\bm p}$. If we, however, keep in mind that the
hard collisions are treated as an average friction force acting on all
ions and electrons, we get a Vlasov-like equation of the form
\beq
\frac{\partial f_e({\bm p},t)}{\partial t}
  - e \left[{\bm E}_{ext}(t)
  - \nabla\Phi
  - \frac{R}{m_e} {\bm V}
    \right] \frac{\partial f_e({\bm p},t)}{\partial {\bm p}}
  = 0 \,,
\label{eq:vlasov}
\eeq
which is consistent with the current balance equation \eqref{eq:mot_current}.
Such a treatment implies that all individual electrons feel the same, average
friction force as the electron fluid.

Next a frame where the electron distribution can be expanded about its
equilibrium form has to be defined. Due to the incorporation of hard
collisions, it differs from the one of freely quivering electrons and is
determined by the external field and the friction force (generalized
Kramers-Hennenberger frame). The transformation to this frame is given by
\begin{subequations}
\label{eq:transform}
\bea
{\bm \rho} \!&=&\! {\bm r} + \gamma {\bm \epsilon}
                   \left[ \sin(\omega_0 t)
                       + \frac{\nuhc}{\omega_0} \cos {(\omega_0 t)} \right] \,,
\\
{\bm u}    \!&=&\! {\bm v} + \omega_0\gamma{\bm \epsilon}
                   \left[ \cos(\omega_0 t)
                        - \frac{\nuhc}{\omega_0} \sin {(\omega_0 t)} \right] \,,
\eea
\end{subequations}
where the abbreviation ${\bm \epsilon} \!=\! -e{\bm E_0}/m\omega_0^2$ is used.
In this frame, the electron distribution can be linearized about its
equilibrium $f_0$ which yields
\beq
\frac{\partial f_1(m_e {\bm u},t)}{\partial t}
   + \frac{e}{m} \frac{\partial \Phi_1({\bf \rho},t)}{\partial {\bm \rho}}
     \cdot \frac{\partial f_0(m_e {\bm u},t)}{\partial {\bm u}} = 0 \,.
\label{eq:small-f}
\eeq
The equation for the lineraised potential $\Phi_1$ is easily obtained from the
Poisson equation \eqref{eq:Poisson} using the first order electron fluctuation
$f_1$ and a coordinate translation according to \eqref{eq:transform}. After
Fourier transformation, the potential can be written in the form 
$\Phi_1({\bm k}, \omega) \!=\! \Sigma({\bm k},\omega)/\varepsilon(k,\omega)$
where $\varepsilon(k,\omega)$ is the dielectric function. The background
source term $\Sigma({\bm k},\omega)$ can be expanded using Bessel functions of the
first kind to yield 
\bea
\Sigma({\bm k},\omega) \!&=&\! \sum_{n,m}^{\infty}
                          \frac{Z e(-1)^n(-i)^m}{2\pi^2k^2} \:
                          J_n(\xi) \, J_m(\bar{\xi})
\nonumber \\
                    &~&\! \times \delta\bigl(\omega + [n+m] \omega_0 \bigr)
                                 \, \sum_j e^{-i{\bm k}\cdot{\bm r}_j} \,,
\label{eq:background}
\eea
where the arguments of the Bessel-functions are given by
$\xi \!=\! \gamma \keps$ and $\bar{\xi} \!=\! \gamma \nuhc \keps / \omega_0$.
The time-dependent fields follow by inverse Fourier transformation and one
obtains ${\bm P}({\bm k},t ) \!=\! i {\bm k} \Phi_1({\bm k},t)$ for the
polarization field. This field is now transformed back to the rest frame of the
ions and averaged over ion positions. In the last step, the sum over
exponentials in Eq.~\eqref{eq:background} becomes the static ion-ion structure
factor $S_{ii}(k)$. For the polarization in coordinate space, one gets
\bea
\label{eq:e_rt}
{\bm P}({\bm r},t ) \!&=&\! i \int d {\bm k} \: {\bm k} \: \phi({\bm k},t)
                                         \, S_{ii}(k)
\nonumber \\
                    \!&~&\! \times \exp\Bigl( i \xi
                                \bigl[ \sin(\omega_0 t) +
                                        \frac{\nuhc}{\omega_0} \cos(\omega_0 t)
                                \bigr]
                                       \Bigr) \,.
\eea
This field can be easily averaged over a laser cycle and then inserted into
the expression (\ref{eq:frequency}) for the collision frequency which yields
\bea
\nu_{ei}(\omega) \!&=&\!
                 - \;\alpha \!\!\!\!
                   \sum_{m,n,s=-\infty}^\infty
                   \int\limits_0^\infty \frac{d \bm k}{k^2} \,
                   \frac{J_n(\xi) \, J_m(\bar{\xi}) \, J_{m+s}(\bar{\xi})}
                        {\varepsilon\bigl(k,[n+m]\omega_0\bigr)} \, S_{ii}(k)
\nonumber \\
              &~&\! \times i^s
        \Bigl[ (n-s)\left(i - \frac{\nuhc}{\omega_0}\right) J_{n-s}(\xi)
               + \bar{\xi} J_{n-s-1}(\xi) \Bigr]
\nonumber \\
              &~&\! + \gamma\nuhc \,,
\label{eq:coll_freq}
\eea
where $\alpha \!=\! (\omega_0 \omega_p / 2\pi^2)(Ze^2/m_e v_0^2)$. Clearly, the
known limiting cases can be readily retrieved. The last term in line 3 
dominates for small laser frequencies giving a Drude-like expression. The
results from quantum Vlasov and Born approximation follow in the weak coupling
limit with $\nuhc \!\to\! 0$. The result of Decker {\em et al.} 
\cite{DMDK_94} follows also from expression \eqref{eq:coll_freq} for
$\nuhc \!\to\! 0$ if the classical dielectric function is used. In this case,
the integral must be truncated at $k_{max}$ to avoid the divergence at small
impact parameters.

The form of expression \eqref{eq:coll_freq} should also clarify why the
collision term was split into a weak coupling and a hard collision part:
only a Drude-like term would have been obtained if the total electron-ion
collision integral was evaluated by using its relation to the stopping power.
Moreover, nonlinear contributions associated with the higher order Bessel
functions would be neglected. Here, the full dynamics in the leading order term
is kept and the effects of strong collisions are incorporated as well.

The derivation above was sketched classically, but can be done
quantum-mechanically as well. The only difference that arises is the form of
the dielectric function $\varepsilon(k,\omega)$ which will be classical or
quantum RPA (compare Ref.~\cite{DMDK_94} with Refs.~\cite{BSHK_01,KP_01}). To
obtain the results shown below in Fig.~\ref{fig:nu_vs_Gamma}, we used the
quantum form including degeneracy corrections for $\varepsilon(k,\omega)$. The
stopping power, that defines the frequency of hard collisions $\nuhc$, is also
obtained from a quantum kinetic descriptions. As a results all integrals can be
performed to infinity and no {\em ad hoc} cutoffs must be introduced.

The main difference of expression \eqref{eq:coll_freq} to other results for
the collision frequency is the incorporation of hard electron-ion collisions.
This has been done by introducing a general friction forces related to the
stopping power of the ions in an electron gas. Many models for the
stopping power have been developed \cite{ZTR_99}, few include hard collisions.
Within quantum statistical theory, they can be described by a T-matrix
approach based on the quantum Boltzmann equation. The related cross sections
are then calculated by numerical solutions of the Schr\"odinger equation
\cite{GS_99}. Dynamic screening effects can be added applying the Gould-DeWitt
scheme \cite{GSK_96} or by velocity-dependent screening length \cite{G_02}.
While both methods coincide for weak and intermediate coupling, only the
latter agrees with simulation data for very high beam-plasma coupling
\cite{G_02,GS_03}. The low velocity part of this model can be expressed in
an analytic fit of the form
\beq
\log(\nu {\rm [th.u.]}) = \exp(-0.0735 x^2 + 1.3373 x - 1.8511) \,,
\eeq
where $x \!=\! Z\Gamma^{3/2}$ and the total collision frequency $\nu$ is given
in thermal units. To obtain the contribution of hard electron-ion collisions
needed in expression \eqref{eq:coll_freq}, the simple Lenard-Balescu form must
be subtracted.

\begin{figure}[t]
\includegraphics[width=8.2cm,clip=true, trim=0mm 0mm 0mm 2.5mm]{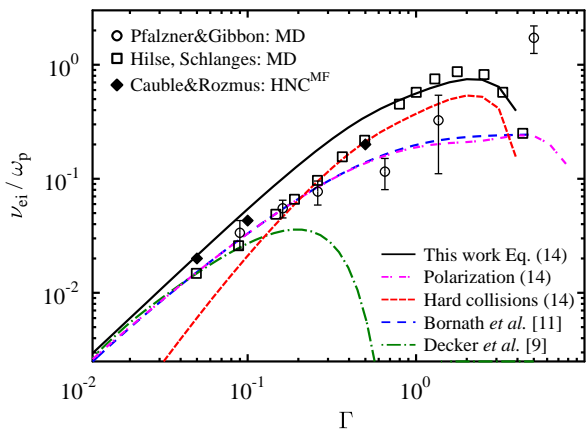}
\caption{Collision frequency $\nu_{ei}$ versus coupling parameter $\Gamma$
         for a hydrogen plasma with $n_e \!=\! 10^{22}\,$cm$^{-3}$ and
         a laser field with $\omega_0/\omega_p \!=\! 3$ and
         $v_0/\vth \!=\! 0.2$.
         Solid line: Eq.~\eqref{eq:coll_freq}; punctured lines: contributions
         of hard collisions and polarization to Eq.~\eqref{eq:coll_freq};
         dashed line: quantum results as in Refs.~\cite{BSHK_01,KP_01};
         dash-dotted line: classical results of Decker {\em et al.}
         \cite{DMDK_94}) with an integral cut off at
         $k_{max} \!=\! m_e \vth^2 / Ze^2$;
         symbols: results of numerical simulations.}
\label{fig:nu_vs_Gamma}
\end{figure}

In Fig.~\ref{fig:nu_vs_Gamma}, the results of our new approach are compared to
other theories and simulation data  \cite{HSBK_05,PG_98,CR_85}. According to
the conditions considered, the stopping power was here calculated using the
combined model of Refs.~\cite{GSK_96,GS_99}, {\em i.e.}, $\nuhc$ is given by
the difference of the T-matrix and the static Born terms. The ions can be
treated individually since $S_{ii} \!\approx\! 1$ even for the highest coupling
strength considered. As expected, all theories agree for weakly coupled
plasmas, but large deviation occur for strong coupling. The classical
description is clearly not applicable here (see Ref.~\cite{GG_09b} for
a discussion of other cutoffs). For a coupling strength of
$\Gamma \!\approx\! 1$, the quantum theories of Refs.~\cite{BSHK_01,KP_01}
also start to disagree with the simulation data. This behavior can be traced
back to i) the neglect of hard collisions and ii) the use of a freely moving
reference frame. As both shortcomings are overcome in our approach we find
excellent agreement with data from MD simulations up to high coupling strength.
Plotting both contributions of Eq.~\eqref{eq:coll_freq} separately reveals that
the Drude term (hard collisions) dominates for high coupling strengths and
defines here the shape of the curve. This adds another nonlinearity to the
result for more highly charged ions since the relevant stopping power is here
not proportional to $Z^2$ \cite{GS_99}.

In conclusion, a new approach for collisional absorption of laser light in
dense plasmas was presented. It incorporates the dynamic response and hard
electron-ion collisions. The latter were introduced by a newly established
connections to the stopping power which also allowed one to define a reference
frame that includes both field effects and friction. Although only results for
the quasi-linear regime $v_0/\vth \!\leq 1\!$ were presented, this approach is
not limited by this requirement and can be easily extended to higher field
amplitudes, correlated ions, and multiple ionization stages.

The authors thank J.~Vorberger (CFSA, Warwick) for fruitful discussions and
EPSRC for financial support.




\begin{thebibliography}{10}
\bibitem{Lindl_04}J.D.~Lindl {\em et al.},
                  Phys.~Plasmas {\bf 11}, 339 (2004).
\bibitem{MDWW_09} M.~Michel {\em et al.},
                  Phys.~Rev.~Lett.~{\bf 102}, 025004 (2009).
\bibitem{K_07}    M.H.~Key,
                  Phys.~Plasmas {\bf 14}, 055502 (2007).
\bibitem{KNCD_08} A.L.~Kritcher {\em et al.},
                  Science {\bf 322}, 69 (2008).
\bibitem{GGGV_08} E.~Garc\'ia Saiz {\em et al.},
                  Nature Phys.~{\bf 4}, 940 (2008).
\bibitem{MHBK_08} B.~F.~Murphy {\em et al.},
                  Phys.~Rev.~Lett.~{\bf 101}, 203401 (2008).
\bibitem{BHS_07}  Th.~Bornath {\em et al.},
                  Laser Physics {\bf 17}, 591 (2007).
\bibitem{DO_62}   J.M.~Dawson, C.~Oberman,
                  Phys.~Fluids~{\bf 5}, 517 (1962).
\bibitem{DMDK_94} C.D.~Decker {\em et al.},
                  Phys.~Plasmas {\bf 1}, 4043 (1994).
\bibitem{KBBS_99} D.~Kremp {\em et al.},
                  Phys.~Rev.~E {\bf 60}, 4725 (1999).
\bibitem{BSHK_01} Th.~Bornath {\em et al.},
                  Phys.~Rev.~E {\bf 64}, 026414 (2001).
\bibitem{KP_01}   H.J.~Kull and L.~Plagne,
                  Phys.~Plasmas~{\bf 8}, 5244 (2001).
\bibitem{RRRW_00} H.~Reinholz {\em et al.},
                  Phys.~Rev.~E {\bf 62}, 5648 (2000).
\bibitem{WMRR_01} A.~Wierling {\em et al.},
                  Phys.~Plasmas {\bf 8}, 3810 (2001).
\bibitem{GS_99}   D.O.~Gericke and M.~Schlanges,
                  Phys.~Rev.~E {\bf 60}, 904 (1999).
\bibitem{G_02}    D.O.~Gericke,
                  Laser \& Part Beams {\bf 20}, 471 (2002).
\bibitem{A_78}    N.R.~Arista,
                  Phys.~Rev.~B {\bf 18}, 1 (1978).
\bibitem{ZD_97}   G.~Zwicknagel and C.~Deutsch,
                  Phys.~Rev.~E {\bf 56}, 970 (1997).
\bibitem{PG_98}   S.~Pfalzner and P.~Gibbon,
                  Phys.~Rev.~E {\bf 57}, 4698 (1998).
\bibitem{HSBK_05} P.~Hilse {\em et al.},
                  Phys.~Rev.~E {\bf 71}, 056408 (2005).
\bibitem{S_65}    V.P.~Silin,
                  Sov.~Phys.~JETP {\bf 20}, 1510 (1965).
\bibitem{GG_09b}  A.~Grinenko and D.O.~Gericke,
                  to be published.
\bibitem{CR_85}   R.~Cauble and W.~Rozmus,
                  Phys.~Fluids {\bf 28}, 3387 (1985).
\bibitem{GSK_96}  D.O.~Gericke {\em et al.},
                  Phys.~Lett.~A {\bf 222}, 241 (1996).
\bibitem{ZTR_99}  G.\ Zwicknagel {\em et al.},
                  Phys.\ Rep.\ {\bf 309}, 117 (1999).
\bibitem{GS_03}   D.O.~Gericke and M.~Schlanges,
                  Phys.~Rev.~E {\bf 67}, 037401 (2003).
\bibitem{GG_09a}  A.~Grinenko and D.O.~Gericke,
                  J.~Phys.~A (accepted).
\end{thebibliography}
\end{document}